\pdfoutput=1
\documentclass[11pt,twoside,a4paper,cmspaper,final,collab]{cms-tdr}
\def\svnVersion{e5734ac-D}\def\svnDate{2020/12/23}\def\cmsCernNoTag{CERN-EP-2020-106}\def\cmsCernDate{\today}\def\cmsMessage{"Published in the Journal of High Energy Physics as \href{http://dx.doi.org/10.1007/JHEP03(2021)003}{\doi{10.1007/JHEP03(2021)003}.}"}
\begin{document}\cmsNoteHeader{HIG-19-002}

\providecommand{\cmsTable}[1]{\resizebox{\textwidth}{!}{#1}}
\newcommand{\hww}{\ensuremath{\PH\to\PWp\PWm}\xspace}
\newcommand{\enmn}{\ensuremath{\Pepm\Pgm^{\mp}\PGn\PAGn}\xspace}
\newcommand{\hwwenmn}{\ensuremath{\PH\to\PWp\PWm\to\enmn}\xspace}
\newcommand{\ww}{\ensuremath{\PWp\PWm}\xspace}
\newcommand{\tautau}{\ensuremath{\PGtp{}\PGtm}\xspace}
\newcommand{\ttst}{\ensuremath{\ttbar{}+{}\PQt\PW}\xspace}
\newcommand{\ttH}{\ensuremath{\ttbar{}\PH}\xspace}
\newcommand{\VH}{\ensuremath{\PV{}\PH}\xspace}
\newcommand{\wj}{\ensuremath{\PW{}+{}\text{jets}}\xspace}
\newcommand{\wg}{\ensuremath{\PW{}\Pgg}\xspace}
\newcommand{\wgs}{\ensuremath{\PW{}\Pgg^{*}}\xspace}
\newcommand{\zg}{\ensuremath{\PZ{}\Pgg}\xspace}
\newcommand{\fbns}{\mbox{\ensuremath{\text{fb}}}\xspace}
\newcommand{\mui}{\ensuremath{\mu_{i}}\xspace}
\newcommand{\mufid}{\ensuremath{\mu^{\text{fid}}}\xspace}
\newcommand{\sigmai}{\ensuremath{\sigma_{i}}\xspace}
\newcommand{\sobs}{\ensuremath{\sigma^{\text{obs}}}\xspace}
\newcommand{\sobsi}{\ensuremath{\sobs_{i}}\xspace}
\newcommand{\ssm}{\ensuremath{\sigma^{\text{SM}}}\xspace}
\newcommand{\ssmi}{\ensuremath{\ssm_{i}}\xspace}
\newcommand{\sfid}{\ensuremath{\sigma^{\text{fid}}}\xspace}
\newcommand{\ptH}{\ensuremath{\pt^{\PH}}\xspace}
\newcommand{\njet}{\ensuremath{N_{\text{jet}}}\xspace}
\newcommand{\mTH}{\ensuremath{\mT^{\PH}}\xspace}
\newcommand{\mTltwo}{\ensuremath{\mT^{l_{2}}}\xspace}
\newcommand{\mll}{\ensuremath{m^{ll}}\xspace}
\newcommand{\ptll}{\ensuremath{\pt^{ll}}\xspace}
\newcommand{\ptlone}{\ensuremath{\pt^{l_{1}}}\xspace}
\newcommand{\ptltwo}{\ensuremath{\pt^{l_{2}}}\xspace}
\newcommand{\ptvecll}{\ensuremath{\ptvec^{ll}}\xspace}
\newcommand{\ptvecltwo}{\ensuremath{\ptvec^{l_{2}}}\xspace}
\newcommand{\mllmtH}{\ensuremath{(\mll, \mTH)}\xspace}
\newcommand{\fourway}{\ensuremath{4\mathcal{W}}\xspace}
\newcommand{\threeway}{\ensuremath{3\mathcal{W}}\xspace}
\newcommand{\twoway}{\ensuremath{2\mathcal{W}}\xspace}
\newcommand{\oneway}{\ensuremath{1\mathcal{W}}\xspace}
\newcommand{\systexp}{\ensuremath{\,\text{(exp)}}\xspace}
\newcommand{\systsignal}{\ensuremath{\,\text{(signal)}}\xspace}
\newcommand{\systbkg}{\ensuremath{\,\text{(bkg)}}\xspace}
\newcommand{\diffbasis}{\ensuremath{\mathcal{DO}}\xspace}
\newcommand{\recolevel}{\ensuremath{\mathcal{RL}}\xspace}
\newcommand{\genlevel}{\ensuremath{\mathcal{GL}}\xspace}

\cmsNoteHeader{HIG-19-002}

\title{Measurement of the inclusive and differential Higgs boson production cross sections in the leptonic {\PW{}\PW} decay mode at $\sqrt{s} = 13\TeV$}

\author{The CMS Collaboration}

\date{\today}

\abstract{
  Measurements of the fiducial inclusive and differential production cross sections of the Higgs boson in proton-proton collisions
  at $\sqrt{s} = 13\TeV$ are performed using events where the Higgs
  boson decays into a pair of {\PW} bosons that subsequently decay into a final state with an
  electron, a muon, and a pair of neutrinos. The analysis is based on data
  collected with the CMS detector at the LHC during 2016--2018, corresponding to an integrated
  luminosity of 137\fbinv. Production cross sections are measured
  as a function of the transverse momentum of the Higgs boson and the
  associated jet multiplicity. The Higgs boson signal is extracted and
  simultaneously unfolded to correct for selection efficiency and resolution
  effects using maximum-likelihood fits to the observed distributions in data.
  The integrated fiducial cross section is measured to be $86.5\pm 9.5\unit{fb}$, consistent with the Standard Model expectation of $82.5\pm 4.2\unit{fb}$.
  No significant deviation from the Standard Model expectations is observed in the differential measurements.
}

\hypersetup{%
pdfauthor={CMS Collaboration},%
pdftitle={Measurement of the inclusive and differential Higgs boson production cross sections in the leptonic WW decay mode at sqrt(s) = 13 TeV},%
pdfsubject={CMS},%
pdfkeywords={CMS, physics, Higgs boson, differential cross section, unfolding}}

\maketitle

\section{Introduction}
\label{sec:introduction}

The Higgs boson, observed by the ATLAS and CMS experiments~\cite{Aad:2012tfa,Chatrchyan:2012xdj,Chatrchyan:2013lba},
has a rich set of properties whose measurements will have a significant impact on the understanding of the physics of the standard model (SM) and possible extensions beyond the
SM (BSM). Extensive effort has been dedicated to determine its quantum numbers and
couplings with ever-improving accuracy due to the large data sample delivered by the CERN LHC and innovations in analysis techniques.

The differential production cross sections of the Higgs boson can be predicted with high precision
and can therefore provide a useful probe of the effects from higher-order corrections in perturbative
theory or any deviation of its properties from the SM expectations. In particular, the differential cross section
as a function of the transverse momentum of the Higgs boson (\ptH) is computed up to
next-to-next-to-leading order (NNLO) in quantum chromodynamics (QCD)~\cite{Monni:2019whf,Jones:2018hbb,Hamilton:2015nsa,Hamilton:2013fea,Cacciari:2015jma,Alioli:2019qzz}, and is known to be
sensitive to possible deviations from the SM in the Yukawa couplings of light
quarks~\cite{Bishara:2016jga} and to effective operators of dimension six or higher in
BSM Lagrangians~\cite{Grazzini:2016paz}.

We present measurements of differential cross sections for Higgs boson production in proton-proton ({\Pp{}\Pp}) collisions at $\sqrt{s} = 13\TeV$
within a fiducial region, as a function of \ptH and jet multiplicity (\njet). 
These two observables are collectively referred to as differential-basis observables (\diffbasis)
hereafter. 
The measurements include all Higgs boson production
modes.
Higgs bosons decaying to two {\PW} bosons that subsequently decay
leptonically into the \enmn final state are considered. 
The data in
these measurements were recorded at the CMS experiment and  correspond to an integrated luminosity of
137\fbinv.

Inclusive Higgs boson production cross sections in the \hww decay mode have been performed by both
ATLAS and CMS~\cite{Aaboud:2018jqu,Sirunyan:2018egh} at $\sqrt{s}=13\TeV$ with smaller data samples.
Both experiments have also reported measurements of differential production cross
sections of the Higgs boson with smaller data samples~\cite{Aaboud:2018ezd,Sirunyan:2018sgc}. In particular, the CMS
Collaboration has measured cross sections as a function of several observables, including \ptH and \njet, using Higgs bosons decaying
into pairs of photons~\cite{Sirunyan:2018kta} and {\PZ} bosons~\cite{Sirunyan:2017exp} at $\sqrt{s} = 13\TeV$
in 35.9\fbinv of data. These measurements have been combined~\cite{Sirunyan:2018sgc}, including in the \ptH spectra data from the search for the Higgs boson produced with large \pt and decaying to a bottom quark-antiquark pair~\cite{Sirunyan:2017dgc}.
The larger branching ratio makes the \enmn final state competitive with the two-photon and two-{\PZ} boson channels.  
Additionally, unlike the decay channel into a bottom quark-antiquark pair, identification of Higgs boson production events in
the \enmn final state does not require the Higgs boson to be boosted, allowing the full range of \ptH to be studied.
In the \hww channel, previous measurements of the differential cross sections
were reported in data collected at $\sqrt{s} = 8\TeV$~\cite{Aad:2016lvc,Khachatryan:2016vnn}. 
Measurements reported in this paper have been performed for the first time in the \hww decay channel at $\sqrt{s} = 13\TeV$, exploiting the full data sample available. 
The methods for the determination of the differential cross section have been updated substantially compared to the 8\TeV measurement~\cite{Khachatryan:2016vnn}, combining the signal extraction, unfolding, and regularization into a single simultaneous fit.

\section{The CMS detector and object selection}
\label{sec:detector}

The central feature of the CMS apparatus is a superconducting solenoid of 6\unit{m} internal
diameter, providing a magnetic field of 3.8\unit{T}. Within the solenoid volume are a silicon pixel
and strip tracker, a lead tungstate crystal electromagnetic calorimeter (ECAL), and a brass and
scintillator hadron calorimeter (HCAL), each composed of a barrel and two endcap sections. Forward
calorimeters extend the pseudorapidity ($\eta$) coverage provided by the barrel
and endcap detectors. Muons are detected using three technologies: drift tubes,
cathode strip chambers, and resistive-plate chambers embedded in the steel
flux-return yoke outside the solenoid.
The muon detectors cover the full $2\pi$ of azimuth ($\phi$) about the beam axis and a range of $\abs{\eta} <2.4$.

Events of interest are selected using a two-tiered trigger system~\cite{Khachatryan:2016bia}. The
first level (L1), composed of specialized hardware processors, uses information from the calorimeters and
muon detectors to select events at a rate of $\approx$100\,kHz within a fixed time interval of
4\mus. The second level, known as the high-level trigger (HLT), consists of a farm of
processors running a version of the full event reconstruction software optimized for fast
processing, and reduces the event rate to $\approx$1\,kHz before data storage.

A more detailed description of the CMS detector, together with a definition of the coordinate system
and the kinematic variables, can be found in Ref.~\cite{Chatrchyan:2008zzk}.

Electrons are identified and their momentum measured in the pseudorapidity interval $\abs{\eta} < 2.5$ by combining the energy measurement in the ECAL, the momentum
measurement in the tracker and the energy sum of all bremsstrahlung photons spatially compatible with
originating from the electron track. 
The single electron trigger efficiency exceeds  90\%  over  the  full $\eta$ range, the efficiency to reconstruct and identify electrons ranges between 60 and 80\% depending on the lepton \pt. 
The momentum resolution for electrons with $\pt\approx 45\GeV$ from $\PZ \to \Pe{}\Pe$ decays ranges from 1.7 to 4.5\% depending on the $\eta$ region. The resolution is
generally better in the barrel than in the endcaps and also depends on the bremsstrahlung
energy emitted by the electron as it traverses the material in front of the
ECAL~\cite{Khachatryan:2015hwa}. 

Muons are identified and their momentum measured in the pseudorapidity interval $\abs{\eta} < 2.4$ matching tracks in the muon chambers and in the silicon tracker.
The single muon trigger efficiency exceeds  90\%  over  the  full $\eta$ range,  and  the  efficiency  to reconstruct and identify muons is greater than 96\%.
The relative transverse momentum resolution for muons with \pt up to 100\GeV is 1\% in
the barrel and 3\% in the endcaps~\cite{Sirunyan:2018fpa,Sirunyan:2019yvv}.

Proton-proton interaction vertices are reconstructed from tracks using the Adaptive Vertex Fitting
algorithm~\cite{Fruhwirth:2007hz}. The candidate vertex with the largest value of summed
physics-object $\pt^2$ is taken to be the primary $\Pp\Pp$ interaction vertex. The physics objects
are the track-only jets, clustered using the jet finding algorithm~\cite{Cacciari:2008gp,Cacciari:2011ma} with
the tracks assigned to candidate vertices as inputs, and the associated missing transverse momentum,
taken as the negative vector sum of the \pt of those jets.

The particle-flow (PF) algorithm~\cite{CMS-PRF-14-001} aims to reconstruct and identify each individual
particle in an event, with an optimized combination of information from the various elements of the
CMS detector. The momenta of electrons and muons are obtained as described above. The energies of
photons are based on the measurement in the ECAL. The energies of charged hadrons are determined from a
combination of their momenta measured in the tracker and the matching ECAL and HCAL energy
deposits.
Finally, the energies of neutral hadrons are obtained from their corresponding 
corrected ECAL and HCAL energies. Such reconstructed particle candidates are generically referred to
as PF candidates.

The hadronic jets in each event are clustered from the PF candidates using the
anti-\kt algorithm~\cite{Cacciari:2008gp, Cacciari:2011ma} with a distance parameter of 0.4. The jet momentum is determined from the vectorial sum
of all particle momenta in the jet. From simulation, reconstructed jet momentum is found to be, on
average, within 5 to 10\% of the momentum of generator jets, which are jets clustered from all generator
final-state particles excluding neutrinos, over the entire \pt spectrum and detector
acceptance. Additional \Pp{}\Pp interactions within the same or nearby bunch crossings (pileup)
can contribute additional tracks and calorimetric energy deposits to the jet momentum. To
mitigate this effect, charged particles identified as originating from pileup vertices are
discarded and an offset correction is applied to correct for remaining contributions from neutral
pileup particles~\cite{CMS-PRF-14-001}. Jet energy corrections are derived from simulation studies
so that the average measured response of jets becomes identical to that of generator jets.  In
situ measurements of the momentum imbalance in dijet, $\text{photon}+\text{jet}$, $\PZ+
\text{jet}$, and multijet events are used to account for any residual differences in jet energy
scale in data and simulation~\cite{Khachatryan:2016kdb,CMS-DP-2020-019}. The jet energy resolution amounts typically
to 15\% at 10\GeV, 8\% at 100\GeV, and 4\% at 1\TeV. Additional selection criteria are applied to
each jet to remove jets potentially dominated by anomalous contributions from various subdetector
components or reconstruction failures. Jets are measured in the range $\abs{\eta}
<4.7$. In the analysis of data recorded in 2017, to eliminate spurious jets caused by detector noise,
all jets were excluded in the range $2.5<\abs{\eta}<3.0$.

The identification of jets containing hadrons with bottom quarks is referred to as {\cPqb} tagging. For
each reconstructed jet, a {\cPqb} tagging score is calculated through a multivariate analysis of jet properties
based on a boosted decision tree algorithm and deep neural networks~\cite{Sirunyan:2017ezt}. Jets
are considered {\cPqb} tagged if this score is above a threshold set to achieve $\approx$80\%
efficiency for bottom-quark jets in \ttbar events. For this threshold, the probability of misidentifying 
charm-quark and light-flavor jets produced in \ttbar events as bottom-quark jets is $\approx$6\%.

Missing transverse momentum (\ptvecmiss) is defined as the negative vector sum of the transverse
momenta of all the PF candidates in an event~\cite{Sirunyan:2019kia}, weighted by their estimated
probability to originate from the primary interaction vertex. The pileup-per-particle identification algorithm~\cite{Bertolini:2014bba} is employed to calculate this probability.

\section{Data sets and simulated samples}
\label{sec:datasets}

The analyzed data sets were recorded in 2016, 2017, and 2018, with corresponding integrated luminosities of
35.9, 41.5, and 59.7\fbinv, respectively~\cite{CMS-PAS-LUM-17-001,CMS-PAS-LUM-17-004,CMS-PAS-LUM-18-002}.

The events in this analysis are selected through HLT algorithms that require the presence of either
a single high-\pt lepton or both an electron and a muon at lower \pt thresholds that pass identification
and isolation requirements. The requirements in the single-lepton triggers are more restrictive than
in the electron-muon triggers, but are less stringent than those applied in the event-selection
stage. In the 2016 data set, the \pt threshold of the single-electron trigger is 25\GeV for $\abs{\eta}
< 2.1$ and 27\GeV for $2.1 < \abs{\eta} < 2.5$, although the use of tight L1 \pt constraints at the beginning of the fill made the effective thresholds higher. The threshold for the single-muon trigger is
24\GeV for $\abs{\eta} < 2.4$. The \pt thresholds in the dilepton trigger are respectively 23
and 8\GeV for the leading and trailing (second highest \pt) leptons for the first part of the data set corresponding to
an integrated luminosity of 17.7\fbinv. The threshold for the trailing lepton is raised to
12\GeV in the later part of the 2016 data set. In the 2017 data set, single-electron and single-muon \pt thresholds are raised to 35
and 27\GeV, respectively. The corresponding thresholds in the 2018 data set are 32 and
24\GeV. The dilepton triggers in the 2017 and 2018 data sets have the same thresholds as given above
for the latter part of the 2016 data set.

Monte Carlo (MC) simulated events are used in this analysis for signal modeling and
background estimation. To account for changes in detector and pileup conditions and to incorporate the
latest updates of the reconstruction software, a different simulation is used in the analysis of each of
the 2016, 2017, and 2018 data sets.  Different event generators are used depending on the simulated
hard scattering processes, but parton distribution functions (PDFs) and underlying event (UE) tunes are
common to all simulated events for a given data set.  The parton-showering and
hadronization processes are simulated through \PYTHIA~\cite{Sjostrand:2014zea} 8.226 (8.230) in 2016
(2017 and 2018). The PDF set is NNPDF 3.0~\cite{Ball:2013hta,Ball:2011uy} (3.1~\cite{Ball:2017nwa})
and the UE tune is CUETP8M1~\cite{Khachatryan:2015pea} (CP5~\cite{Sirunyan:2019dfx}) for the 2016
sample (2017 and 2018 samples).

Higgs boson production through gluon-gluon fusion (ggF), vector-boson fusion (VBF),
weak-boson associated production (\VH, with {\PV} representing either the {\PW} or {\PZ} boson), and \ttbar
associated production (\ttH), are considered as signal processes in this analysis. Weak boson associated production
has contributions from quark- and gluon-induced {\PZ} boson associated production and {\PW} boson
associated production.  Events for all signal production channels are generated using \POWHEG
v2~\cite{Nason:2004rx,Frixione:2007vw,Alioli:2010xd,Bagnaschi:2011tu,Nason:2009ai,Luisoni:2013kna,Hartanto:2015uka} at next-to-leading order (NLO) accuracy in
QCD, including finite quark mass effects. The ggF events are further
reweighted to match the NNLOPS~\cite{Hamilton:2013fea,Hamilton:2015nsa} prediction  in the distributions
of \ptH and \njet. The reweighting is based on \ptH and \njet as computed in the Higgs boson
simplified template cross section (STXS) scheme 1.0~\cite{Berger:2019wnu}.
All signal samples are normalized to the cross sections recommended in~\cite{deFlorian:2016spz}. In particular, the ggF sample is normalized to next-to-next-to-next-to-leading order (N3LO) QCD accuracy and NLO electroweak accuracy~\cite{Anastasiou:2014lda, Anastasiou:2015yha, Anastasiou:2016cez}. 
Alternative sets of events for ggF and VBF production using the
\MGvATNLO v2.2.2 generator~\cite{Alwall:2014hca} are used for comparison with the extracted
differential cross sections. The alternative ggF sample is generated with up to two extra partons merged through the FxFx scheme~\cite{Frederix:2012ps} in the infinite top quark mass limit.
The Higgs boson mass is assumed to be 125\GeV for these simulations.

The \textsc{JHUgen} generator~\cite{Bolognesi:2012mm} (v5.2.5 and 7.1.4 in 2016 and 2017--2018, respectively) is used to simulate the decay of the Higgs boson into two
{\PW} bosons and subsequently into leptons for the VBF events in 2016, ggF and VBF events from 2017 and
2018, and quark-induced {\PZ{}\PH} production in 2017 and 2018. 
The decay of the
Higgs boson in other signal samples is simulated through \PYTHIA 8.212 along with the parton shower (PS) and
hadronization. Higgs boson that decays into a \tautau pair is considered as background in this analysis.

Quark-initiated nonresonant {\PW} boson pair production (\ww) is simulated at NLO with \POWHEG v2~\cite{Nason:2013ydw}. Gluon-initiated, loop-induced nonresonant \ww is simulated with  \MCFM v7.0~\cite{Campbell:1999ah,Campbell:2011bn,Campbell:2015qma} and normalized to its NLO cross section~\cite{Caola:2016trd}. The \ttbar and single top production (\ttst) are simulated with \POWHEG v2~\cite{Frixione:2007nw,Alioli:2009je,Re:2010bp}. The Drell--Yan \Pgt lepton pair
production (\tautau) is simulated with \MGvATNLO v2.4.2 with up to two additional jets at NLO accuracy. Radiative {\PW} production (\wg) is simulated with \MGvATNLO v2.4.2 with up to 3 additional jets at LO accuracy. Other diboson processes involving at least one {\PZ} boson  or a virtual photon ({\Pgg}$^{*}$) with mass down to 100~\MeV are simulated with \POWHEG v2~\cite{Nason:2013ydw}. Associated \wgs production with virtual photon mass below 100~\MeV is simulated by the parton shower on top of the \wg sample. The \wgs prediction is corrected with a scale factor extracted from a trilepton control region, following the approach described in Ref.~\cite{Sirunyan:2018egh}. Purely electroweak \ww plus two jets production is simulated at LO with \MGvATNLO v2.4.2. Multiboson production with more than two vector bosons is simulated at NLO with \MGvATNLO v2.4.2.

The simulated quark-induced \ww background is weighted event-by-event to match the
transverse momentum distribution of the \ww system to NNLO plus next-to-next-to leading logarithm
(NNLL) accuracy in QCD~\cite{Meade:2014fca,Jaiswal:2014yba}. It is also weighted to include the effect of electroweak corrections, computed based on Ref.~\cite{Gieseke:2014gka}. The \ttbar component of the \ttst
background and the \tautau events are also weighted to improve agreement of the
simulated \pt distributions of the \ttbar and Drell--Yan systems with data~\cite{Khachatryan:2016mnb,Sirunyan:2019bzr}.

For all processes, the detector response is simulated using a detailed description of the CMS
detector, based on the \GEANTfour package~\cite{Agostinelli:2002hh}.  To model multiple {\Pp{}\Pp} collisions in one beam crossing, minimum bias events simulated in \PYTHIA are overlaid onto each event, with the
number of interactions drawn from a distribution that is similar to the observed distribution. The average number of such interactions per event is $\approx$23 for the
2016 data, and 32 for the 2017 and 2018 data.

To mitigate the discrepancies between data and simulation in various distributions, simulated events
are reweighted according to relevant lepton or jet kinematic variables. Discrepancies due to multiple
causes, such as the difference in the pileup distribution and the imperfect modeling of the
detector, are corrected using weights derived from comparisons of simulation with observed data in control regions.

\section{Analysis strategy}
\label{sec:strategy}

The differential production cross sections are measured using dilepton event samples selected based
on the reconstructed properties of the leptons and \ptvecmiss. Events passing the selections described in Section~\ref{sec:selection} are
referred to as signal candidate events, and are split into reconstruction-level (\recolevel) bins of the \diffbasis. The \recolevel \ptH is computed as the
magnitude of the vectorial sum of the transverse momenta of the two lepton candidates and
\ptvecmiss. The missing transverse momentum represents the total vector \pt of the two
neutrinos that escape detection. The \recolevel \njet is the number of jets
with $\pt > 30\GeV$ and $\abs{\eta} < 4.7$.

The signal candidate events are dominated by background processes, with main
contributions from \ww, \ttst, \tautau, and events with misidentified leptons or leptons from
heavy-flavor hadron decays (nonprompt leptons). 
The total number of signal events in the sample is extracted by template fitting techniques, exploiting quantities that separate signal and background.

Two observables, the dilepton mass (\mll) and the transverse mass of the Higgs boson (\mTH),
are found to have strong discrimination power against background processes. The value of \mTH can be
defined as
\begin{equation}
  \mTH = \sqrt{2 \ptll \ptmiss \left[1 - \cos\Delta\phi(\ptvecll, \ptvecmiss)\right]},
\end{equation}
where \ptll is the magnitude of the vector sum of the transverse momenta of the two
lepton candidates, and $\Delta\phi(\ptvecll, \ptvecmiss)$ is the azimuthal angle between
\ptvecll and \ptvecmiss.

Signal candidate events in individual \recolevel bins are therefore sorted into
two-dimensional \mllmtH histograms. The number of Higgs boson production signal events in each
histogram can be inferred by fitting it with a model that consists of a sum of background and
signal templates, obtained from their respective expected distributions. The estimation of the background is described briefly in
Section~\ref{sec:background} and more thoroughly in
Refs.~\cite{Sirunyan:2018egh,Chatrchyan:2013iaa}. Signal expectations are
derived from the simulated event samples described in Section~\ref{sec:datasets}. 
There is only a small dependence of the signal \mllmtH shape on production mode, thus distributions from the four
Higgs boson production modes are combined with their relative normalizations fixed to the SM predictions. 

To extract differential cross section measurements from such fits, signal templates from
different bins of \diffbasis values predicted by the event generator
(generator-level, \genlevel, bins) are individually assigned a priori unconstrained normalization
factors. Initial normalizations of the signal templates are set to the SM expectations. The best fit
normalization factor for the templates of a \genlevel bin $i$ can therefore be interpreted as its
signal strength modifier $\mui = \sobsi / \ssmi$, where \sobsi and \ssmi are the observed and
predicted fiducial cross sections in bin $i$.

Generator-level and \recolevel observable values are not perfectly aligned due to resolution and energy scale effects.
For this reason, signal events from one \genlevel bin $i$ contribute to multiple \recolevel bin
templates, which are all scaled together by \mui. Therefore, by performing one simultaneous fit over
all \recolevel bin histograms, signal strength modifiers of the \genlevel observable
bins can be determined exploiting the full statistical power of the data set. 
This fit extracts the signal and simultaneously unfolds the measured cross sections into the \genlevel bins, correctly propagating the experimental covariance matrix.
The unfolding procedure can be highly sensitive to statistical fluctuations in the
observed distributions, especially for the \ptH measurement, where the contributions
from each \genlevel bin into multiple \recolevel bins are significant. To mitigate this effect, 
a regularization procedure is introduced in the fit for the \ptH measurement to obtain the final
result. More details about the fiducial phase space, the fit, and the regularization
scheme are given in Section~\ref{sec:extraction}.

\section{Event selection}
\label{sec:selection}

The selection of signal candidate events starts with a requirement of at least two charged lepton
candidates, where the two with the highest $\pt$ (leading and trailing lepton candidates)
have tracks associated with the primary vertex, and have opposite charge.
The two leptons must be an electron and a muon to suppress Drell--Yan background.
Charged leptons are required to satisfy the isolation criterion that the scalar
sum of the \pt of PF candidates associated with the primary vertex, exclusive
of the lepton itself, and neutral PF particles in a cone of a radius $\Delta R
= \sqrt{\smash[b]{(\Delta\eta)^2 + (\Delta\phi)^2}}$=0.4 (0.3), where $\phi$ is
the azimuthal angle in radians, centered on the muon (electron) direction is
below a threshold of 15 (6)\% relative to the muon (electron) \pt. To mitigate
the effect of the pileup on this isolation variable, a correction based on the
average energy density in the event~\cite{Cacciari:2007fd} is applied.
Additional requirements on the transverse  and longitudinal impact parameters
with respect to the primary vertex are included. An algorithm based on the
evaluation of the track hits in the first tracker layers is used to reject
electrons arising from photon conversions. 
The transverse momenta of the leading and trailing lepton candidates, $\ptlone$ and $\ptltwo$, must be
greater than 25 and 13\GeV, respectively, so that the electron-muon triggers are efficient. To
ensure high reconstruction efficiencies, only electron candidates with $\abs{\eta} < 2.5$ and muon
candidates with $\abs{\eta} < 2.4$ are considered. Other lepton candidates in the event, if there
are any, must have $\pt < 10\GeV$. 

Signal candidate events must further satisfy $\ptmiss > 20\GeV$ and $\ptll > 30\GeV$ to discriminate
against QCD multijet and {\tautau} backgrounds. 
The contribution from the {\tautau} background, including that from the low-mass Drell--Yan process, is further suppressed by the requirements 
$\mll >12\GeV$, $\mTH > 60\GeV$, and $\mTltwo > 30\GeV$.   
Here the last quantity is defined by
\begin{equation}
  \mTltwo = \sqrt{2 \ptltwo \ptmiss \left[1 - \cos\Delta\phi(\ptvecltwo, \ptvecmiss)\right]},
\end{equation}
where {\ptvecltwo} is the transverse momentum of the trailing lepton, {\ptltwo} is the magnitude, and
$\Delta\phi(\ptvecltwo, \ptvecmiss)$ is the opening azimuthal angle relative to \ptvecmiss. 
This observable stands as a proxy to the mass of the virtual {\PW} boson from the Higgs boson decay. 
As such, the last criterion also limits the contribution from nonprompt lepton background due
to single {\PW} boson production, when the trailing lepton candidate is a misidentified jet and
therefore has little correlation with \ptvecmiss. Finally, to suppress {\ttst} events, the events are
required to have no {\cPqb}-tagged jets with $\pt > 20\GeV$.

The event selection criteria are identical among the three data sets, aside from certain details
such as the definition of {\cPqb} tagging. The efficiencies of the signal candidate selection for
identifying ggF events with {\PW} bosons decaying to leptons are 2.8, 3.6, and 3.6\% for the
2016, 2017, and 2018 data sets, respectively.  The differences in efficiencies arise mainly from the
requirements set on lepton identification and \ptmiss resolution.

Within each \recolevel bin of the \diffbasis, signal candidate events
are categorized by \ptltwo and flavors of the leptons to maximize the
sensitivity to signal. Categories with $\ptltwo < 20\GeV$ receive, in comparison to those with
$\ptltwo > 20\GeV$, more contributions from nonprompt-lepton background but less from \ww and
\ttbar processes, and result in fewer total background events. However, the Higgs boson signal is expected to
contribute evenly to the two \ptltwo regions, providing thereby categories with $\ptltwo < 20\GeV$
with larger signal-to-background ratios. Since nonprompt leptons are more likely to arise from jets 
misidentified as electrons, categorization within the two regions by the flavor of the
leptons helps increase the sensitivity by creating two regions with a different signal-to-background ratio. 
This four-way categorization (\fourway) is applied to reconstructed \diffbasis
bins with a sufficiently large expected number of events. 
For bins with fewer expected events, categorization is reduced to three-way (\threeway, using \ptltwo, and flavor categorization 
for $\ptltwo < 20\GeV$), two-way (\twoway, using just \ptltwo), or none (\oneway). 
In the most sensitive categories, the ratio of expected signal yield to the expected total number of events is $\approx$0.08, and
the ratio of expected signal events to the square root of expected background events is 3.5.

Control regions for {\ttst} and {\tautau} background
processes are used to constrain the estimates of these processes in the simultaneous
fit. The definitions of the two control regions follow that of the signal region closely to make 
the event kinematics similar among the three regions. Specifically, both control regions share all event
selection criteria with the signal region except for the requirements on {\mll}, {\mTH}, {\mTltwo},
and the number of {\cPqb}-tagged jets. The {\ttst} control region instead requires $\mll > 50\GeV$
and at least one {\cPqb}-tagged jet with $\pt > 20\GeV$. If there is another jet in the event with
$\pt > 30\GeV$, the {\cPqb}-tagged jets must also have $\pt > 30\GeV$. There is no constraint on
{\mTH}, and the requirement $\mTltwo > 30\GeV$ is common with the signal region. The {\tautau}
control region requires $40 < \mll < 80\GeV$ and $\mTH < 60\GeV$, and has no constraint on
$\mTltwo$. The restriction of having no {\cPqb}-tagged jets with $\pt > 20\GeV$ is common with the
signal region.

\section{Background modeling}
\label{sec:background}

All background processes, except for that from nonprompt lepton events, are modeled using MC simulation. 
The nonprompt lepton background is modeled by applying weights to events containing lepton
candidates passing less stringent selection criteria than those used in the signal region. These weights,
called fake-lepton factors, are obtained from the probability of a jet being misidentified as a
lepton and the efficiency of correctly reconstructing and identifying a lepton. More
details about this method are given in Ref.~\cite{Sirunyan:2018egh}. 
The validity of this background estimate is checked by comparing
the prediction of the \mllmtH distribution of the nonprompt lepton events to
the observed distribution in a control region with two leptons of the same charge.

Different constraints are applied to the background template normalization, to reflect our knowledge of the cross section of those processes in the model.
First, the normalizations of the templates of the three main 
background processes, \ie, \ww, \ttst, and \tautau, are left unconstrained separately in each \recolevel 
bin. This treatment reflects the belief that precise predictions of these background processes are
essential, but the MC simulation cannot be trusted at extreme values of the observables, especially
large \njet. Their normalizations are therefore determined from the observed data. To help
constrain \ttst and \tautau, control samples enriched in the two processes (see Section~\ref{sec:selection}) are included in the
simultaneous fit. The normalizations of the \ttst and \tautau templates in these control
samples are fit with factors that also scale the respective templates in the fit to the
signal candidate events. The normalization of the \ww template is determined without using specific control
samples, and is mostly constrained by the high \mll region.

Normalizations of the templates for the minor background processes are centered at the SM
expectations and are constrained a priori by their respective systematic uncertainties.
Normalizations of the nonprompt lepton templates are centered at the estimates given by the method
described above. Because the closure of the nonprompt background estimation method depends on the flavor composition of the jets faking the leptons, and since the flavor composition varies among \diffbasis bins, the normalization of the nonprompt background is allowed to
vary independently in each of those bins.

\section{Definition of the fiducial region and extraction of the signal}
\label{sec:extraction}

The fiducial region is defined in Table~\ref{tab:fiducial}, with all quantities evaluated at
generator level after parton showering and hadronization. Leptons are
``dressed'', \ie, momenta of photons radiated by leptons within a cone of $\Delta R =
\sqrt{\smash[b]{(\Delta\eta)^2 + (\Delta\phi)^2}} < 0.1$ are added to the lepton momentum. The fiducial region
definition matches that of the event selection criteria, except for the $\eta$ bound of muons
($\abs{\eta} < 2.4$ in the event selection) and the absence of any direct selection of \ptmiss. Generator level \mTH and \mTltwo employ a generator level \ptvecmiss definition corresponding to the vector sum of all neutrinos in the event.
The expected fiducial cross section and its theoretical uncertainty~\cite{deFlorian:2016spz} computed
for the nominal signal is
\begin{equation}
  \label{eqn:expected_sigma}
  \ssm = 82.5\pm 4.2\unit{fb},
\end{equation}
This cross section is estimated using, for each process, the cross sections recommended in~\cite{deFlorian:2016spz} and estimating the acceptance of the fiducial region from the nominal signal samples.
\begin{table}[tbp]
  \centering
  \topcaption{
    Definition of the fiducial region.
  }
  \label{tab:fiducial}
  \begin{tabular}{lr}
    Observable                             & Condition \\
    \hline
    Lepton origin                          & Direct decay of \hww \\
    Lepton flavors; lepton charge          & \Pe\Pgm (not from \Pgt decay); opposite \\
    Leading lepton \pt                     & $\ptlone > 25\GeV$ \\
    Trailing lepton \pt                    & $\ptltwo > 13\GeV$ \\
    $\abs{\eta}$ of leptons                & $\abs{\eta} < 2.5$ \\
    Dilepton mass                          & $\mll > 12\GeV$ \\
    \pt of the dilepton system             & $\ptll > 30\GeV$ \\
    Transverse mass using trailing lepton  & $\mTltwo > 30\GeV$ \\
    Higgs boson transverse mass            & $\mTH > 60\GeV$ \\
  \end{tabular}
\end{table}

The differential production cross sections for the Higgs boson are inferred from the signal strength
modifiers extracted through a simultaneous fit to all bins and categories of signal candidate events
and two control regions. The systematic uncertainties discussed in
Section~\ref{sec:systematics} are represented by constrained or unconstrained nuisance parameters
that affect the shapes and normalizations of the signal and background templates. The simultaneous
fit maximizes the likelihood function
\begin{equation}
  \mathcal{L}(\boldsymbol{\mu}; \boldsymbol{\theta}) = \prod_{j} \mathrm{Poisson} \left( n_j; s_j(\boldsymbol{\mu}; \boldsymbol{\theta}) + b_j(\boldsymbol{\theta}) \right) \mathcal{N}(\boldsymbol{\theta}) \mathcal{K}(\boldsymbol{\mu}).
\end{equation}
In the formula, $\boldsymbol{\mu}$ and $\boldsymbol{\theta}$ are vectors of the signal strength
modifiers and nuisance parameters, respectively. The expression $\mathrm{Poisson}(n;\lambda)$
represents the Poisson probability of observing $n$ events when expecting $\lambda$, and
$n_{j}$ is the observed number of events in a given bin of the \mllmtH template in any \recolevel category, with index
$j$ running over bins of histograms of signal region categories and control regions for all the
\recolevel \diffbasis bins, and all three data sets.
The signal in the $j$th bin is represented by
\begin{equation}
  \label{eqn:signal_term_original}
  s_j(\boldsymbol{\mu}; \boldsymbol{\theta}) = \sum_{i=1}^{N} \left[ A_{ji}(\boldsymbol{\theta}) \mui L_{j} \sigmai \right],
\end{equation}
where $N$ is the number of \genlevel \diffbasis bins.
The migration 
matrix $A_{ji}$ represents the number of events expected in \recolevel bin $j$ for each \hww
signal event found in the \genlevel bin $i$. The expected number of events in
bin $i$ are expressed as a product of $\mui$, the total integrated luminosity $L_{j}$ (with three possible values corresponding to the three data sets), and the signal cross section
\sigmai. Note that here \sigmai contains both fiducial and nonfiducial components. The total background contribution in bin $j$ is represented by $b_j$. The
factor $\mathcal{N}(\boldsymbol{\theta})$ incorporates a priori constraints on the nuisance
parameters, taken as log-normal distributions for most of the individual $\boldsymbol{\theta}$
elements. Finally, the regularization factor $\mathcal{K}(\boldsymbol{\mu})$, present only in the
\ptH measurement, is constructed as
\begin{equation}
  \mathcal{K}(\boldsymbol{\mu}) = \prod_{i=2}^{N-1} \exp \left( \frac{-\left[(\mu_{i+1} - \mu_{i}) - (\mu_{i} - \mu_{i-1})\right]^2}{2 \delta^2} \right),
\end{equation}
with index $i$ running over \genlevel \diffbasis bins, penalizing thereby large variations among signal strength modifiers of neighboring
bins. The parameter $\delta$ controls the strength of the regularization, and is optimized by
minimizing the mean of the global correlation coefficient~\cite{Schmitt:2012kp} in fits to ``Asimov''
data sets~\cite{Cowan:2010js}. The optimal value of $\delta$ is found to be 2.50. It should be noted that the regularization term acts as a smoothing constraint on the unfolded distribution. Because the distribution of \njet is discrete, regularization was not applied in the \njet fit.

Nonfiducial signal events are scaled together with the
fiducial components, with the distinction between fiducial and nonfiducial parts made only when
translating the extracted signal strength modifiers into fiducial differential cross sections, achieved by multiplying the fiducial cross section in a given \genlevel \diffbasis bin $i$ by the corresponding $\mu_{i}$. This
treatment is chosen because the ratio of nonfiducial to fiducial signal yields expected
in this analysis averages across \diffbasis bin to $\approx$0.2. 
This value is significantly larger than for the diphoton and two {\PZ}
boson decay channels, rendering the scaling of just the
fiducial component unphysical. Nonfiducial signal events appear in the signal region mostly through
the discrepancy between \genlevel 
and \recolevel \ptmiss affecting \mTltwo and \mTH. In addition, for larger values of \njet, the leading Higgs boson production mode is \ttH, which
has more possible $\Pepm{}\Pgm^{\mp}$ final-state configurations where the lepton pair does not
arise from \hww decay. The ratio of nonfiducial over fiducial signal yields is however still affected by the uncertainties on the migration matrix, allowing it to vary postfit with respect to its prefit value.

A \textsc{Rivet}~\cite{Bierlich:2019rhm} implementation of the STXS scheme~\cite{deFlorian:2016spz} is used to compute the \genlevel 
\ptH and \njet observables. For \njet, all final-state particles from the primary interaction, excluding the
products from Higgs boson decay, are clustered using the anti-\kt algorithm with a distance
parameter $R=0.4$, and jets with $\pt > 30\GeV$ are counted regardless of their rapidity.

The binning in both \ptH and \njet is common for the fiducial space and for the reconstructed
events. Bin definitions and categorizations of the reconstructed events within each bin are
summarized in Table~\ref{tab:binning}. The bin widths at lower values of \ptH are dictated by the
reconstruction resolution of \ptmiss that affects the resolution of \ptH. At higher
values, boundaries are chosen so that the expected uncertainties in $\mui$ are less than
unity. The fraction of events reconstructed in the correct \genlevel bin ranges from 52 to 73\% when spanning from the lowest to the highest \ptH bin, and the purity of each \ptH bin, \ie,  the fraction of events in \recolevel bin $i$ that also belong to \genlevel bin $i$, ranges from 48 to 80\%. Corresponding numbers for the \njet measurement are 80 to 92\% and 68 to 95\%, respectively, with the highest jet multiplicity bins representing the lowest bound of these intervals.

\begin{table}[tbp]
  \centering
  \topcaption{
  Binning of the \diffbasis and signal categorizations used in the respective bins.
  }
  \label{tab:binning}
  \begin{tabular}{lcccccc}
  \multicolumn{7}{c}{\ptH} \\
  Binning (GeV): & 0--20 & 20--45 & 45--80 & 80--120 & 120--200 & $>$200 \\
  Categorization: & \fourway & \fourway & \fourway & \threeway & \twoway & \twoway \\
  \end{tabular}
  \begin{tabular}{lccccc}
  \multicolumn{6}{c}{\njet} \\
  Binning: & 0 & 1 & 2 & 3 & $\geq$4 \\
  Categorization: & \fourway & \fourway & \twoway & \oneway & \oneway \\
  \end{tabular}
\end{table}

The values of the signal acceptance per \genlevel \diffbasis bin are shown in Tables~\ref{tab:acc_pth} and ~\ref{tab:acc_njet} for \ptH and \njet respectively.
\begin{table}[tbp]
  \centering
  \topcaption{
  Acceptance of each \genlevel \ptH bin with its theoretical uncertainty.
  }
  \label{tab:acc_pth}
  \begin{tabular}{cccccc}
  \ptH &  \multicolumn{5}{c}{Acceptance per production mode (\%)} \\
  (\GeVns)& All SM & ggF & VBF & \VH & \ttH\\
  \hline
      0--20 & $6.77 \pm 0.36$ & $6.79 \pm 0.37$ & $6.57 \pm 0.13$ & $4.99 \pm 0.12$ & $11.3 \pm 1.1$\\
      20--45 & $6.32 \pm 0.30$ & $6.33 \pm 0.33$ & $6.31 \pm 0.09$ & $4.91 \pm 0.08$ & $11.1 \pm 1.1$\\
      45--80 & $5.94 \pm 0.42$ & $5.91 \pm 0.53$ & $5.86 \pm 0.08$ & $4.81 \pm 0.06$ & $10.9 \pm 1.0$\\
      80--120 & $6.13 \pm 0.47$ & $5.99 \pm 0.73$ & $5.77 \pm 0.08$ & $5.21 \pm 0.07$ & $11.1 \pm 1.1$\\
      120--200 & $6.35 \pm 0.59$ & $5.84 \pm 1.06$ & $5.87 \pm 0.09$ & $5.74 \pm 0.08$ & $11.4 \pm 1.1$\\
      $>$200 & $6.89 \pm 0.73$ & $5.87 \pm 1.47$ & $6.22 \pm 0.15$ & $6.43 \pm 0.17$ & $11.9 \pm 1.1$\\
  \end{tabular}
\end{table}

\begin{table}[!t]
  \centering
  \topcaption{
  Acceptance of each \genlevel \njet bin with its theoretical uncertainty.
  }
  \label{tab:acc_njet}
  \begin{tabular}{cccccc}
   &  \multicolumn{5}{c}{Acceptance per production mode (\%)} \\
  \njet & All SM &  ggF & VBF & \VH & \ttH\\
  \hline
      0 & $6.50 \pm 0.35$ & $6.58 \pm 0.37$ & $6.12 \pm 0.11$ & $4.98 \pm 0.06$ & $12.5 \pm 1.3$\\
      1 & $6.03 \pm 0.64$ & $6.04 \pm 0.76$ & $5.91 \pm 0.08$ & $5.24 \pm 0.07$ & $12.5 \pm 1.2$\\
      2 & $6.36 \pm 0.72$ & $6.08 \pm 1.24$ & $5.99 \pm 0.08$ & $5.44 \pm 0.07$ & $12.5 \pm 1.2$\\
      3 & $7.08 \pm 0.73$ & $6.26 \pm 1.30$ & $6.11 \pm 0.11$ & $5.60 \pm 0.10$ & $11.5 \pm 1.1$\\
      $\geq$ 4 & $7.54 \pm 0.66$ & $6.16 \pm 1.45$ & $6.03 \pm 0.20$ & $5.51 \pm 0.15$ & $10.3 \pm 1.0$\\  
  \end{tabular}
\end{table}

The two-dimensional histograms of \mllmtH in the signal region have different
binnings depending on the expected number of events and statistical uncertainties in the templates. The
finest binning is 10--25, 25--40, 40--50, 50--70, 70--90, and $>$90\GeV in \mll; and
60--80, 80--90, 90--110, 110--130, 130--150, and $>$150\GeV in \mTH.
The coarsest binning, used for the highest \ptH bins, is 10--50 and $>$50\GeV in \mll
and 60--110 and $>$110\GeV in \mTH.

\begin{figure}[btp]
  \centering
  \includegraphics[width=0.8\textwidth]{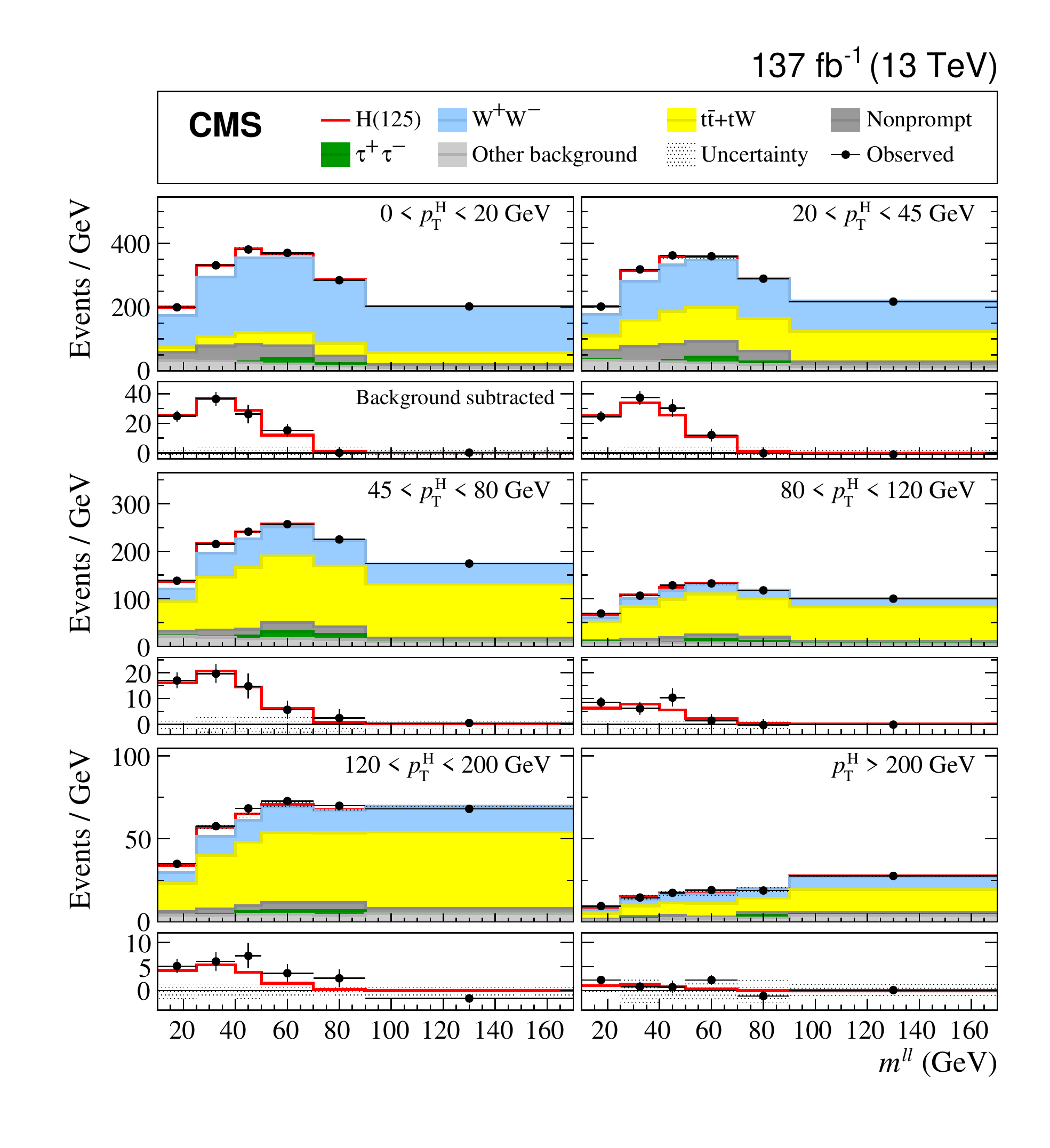}
  \caption{
    Observed distributions of \mll in data and the expectations from the best fit model with the uncertainties. The distributions in each \ptH bin are given in separate panels. Within each panel, the lower sub-panel displays background-subtracted observations and expectations.
  }
  \label{fig:postfit_sr_ptH}
\end{figure}

\begin{figure}[btp]
  \centering
  \includegraphics[width=0.8\textwidth]{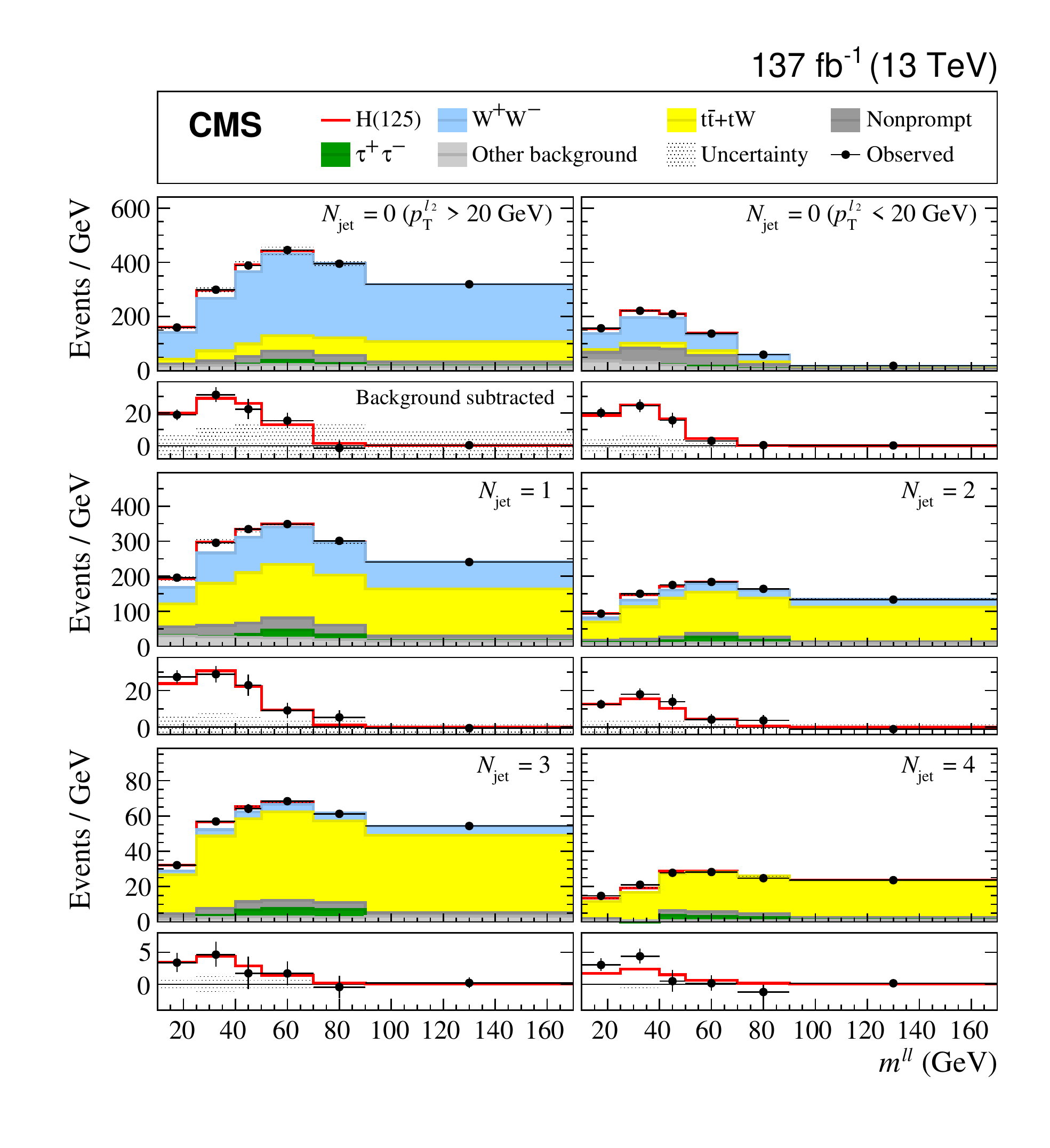}
  \caption{
    Observed distributions of \mll in data and the expectations from the best fit model with the uncertainties. The distributions in each \njet bin are given in separate panels. Within each panel, the lower sub-panel displays background-subtracted observations and expectations.
    For $\njet = 0$, results are split into $\ptltwo > 20\GeV$ (left) and $\ptltwo < 20\GeV$ (right).
  }
  \label{fig:postfit_sr_njet}
\end{figure}

The observed events are shown as a function of \mll in Figs.~\ref{fig:postfit_sr_ptH} and
\ref{fig:postfit_sr_njet}, along with the predictions from the best fit model and their estimated
overall uncertainties. The \mll distributions are formed by integrating the two-dimensional \mllmtH
distributions and templates over \mTH and combining all signal regions and all data sets. The yield breakdown in each \recolevel \diffbasis bin is shown in Tables~\ref{tab:yields_pth} and \ref{tab:yields_njet} for the \ptH and \njet case respectively.

\begin{table}[tbp]
  \centering
  \topcaption{
  Signal and background post-fit (pre-fit) yields in the \recolevel \ptH bins. \label{tab:yields_pth}
  }
\cmsTable{  
\begin{tabular}{c r@{}l@{ }r r@{}l@{ }r r@{}l@{ }r r@{}l@{ }r r@{}l@{ }r r@{}l@{ }r}
          \multirow{2}{*}{Process} & \multicolumn{18}{c}{\recolevel \ptH bin} \\ 
          & \multicolumn{3}{c}{[0--20]} & \multicolumn{3}{c}{[20--45]} & \multicolumn{3}{c}{[45--80]} & \multicolumn{3}{c}{[80--120]} &  \multicolumn{3}{c}{[120--200]} &  \multicolumn{3}{c}{$>$ 200} \\

\hline
           Data   &    \multicolumn{3}{c}{41032}   & \multicolumn{3}{c}{41799} &  \multicolumn{3}{c}{31273}&  \multicolumn{3}{c}{16942} & \multicolumn{3}{c}{10366}  & \multicolumn{3}{c}{3514} \\ [4pt] 
           H(125) &    1485 &$\pm$ 81  &(1356) &    1386 &$\pm$ 80  &(1402) &     835 &$\pm$ 52  &(792) &      320 &$\pm$ 36  &(344) &     217 &$\pm$ 33  &(222)&      54 &$\pm$ 17  &(75) \\ 
 All background   &   39532 &$\pm$ 280 &(35861)&   40414 &$\pm$ 391 &(41978) &  30423 &$\pm$ 393 &(32293) &  16614 &$\pm$ 351 &(17809)&  10154 &$\pm$ 220 &(10790) & 3475 &$\pm$ 107 &(4019) \\ [4pt]

          \tautau &     537 &$\pm$ 49  &(372) &      675 &$\pm$ 43  &(585) &      684 &$\pm$ 61  &(482) &      316 &$\pm$ 42  &(195) &     173 &$\pm$ 24  &(219)   &  104 &$\pm$ 58  &(83) \\
              \ww &   26945 &$\pm$ 213 &(22840) &  17421 &$\pm$ 290 &(18771) &   7444 &$\pm$ 269 &(9048) &    2759 &$\pm$ 250 &(3972) &   2205 &$\pm$ 155 &(2816) &  1037 &$\pm$ 70  &(1637) \\
            \ttst &    5571 &$\pm$ 65  &(5492) &   14700 &$\pm$ 176 &(14528) &  18313 &$\pm$ 239 &(18188) &  11482 &$\pm$ 220 &(11624) &  6481 &$\pm$ 137 &(6488) &  1659 &$\pm$ 40  &(1671) \\
        Nonprompt &    3709 &$\pm$ 127 &(5154) &    4373 &$\pm$ 128 &(5909) &    1822 &$\pm$ 107 &(3143) &    1002 &$\pm$ 80  &(1239) &    558 &$\pm$ 52  &(749) &    197 &$\pm$ 23  &(279) \\
 Other background &    2770 &$\pm$ 102 &(2002) &    3245 &$\pm$ 137 &(2186) &    2160 &$\pm$ 100 &(1431) &    1055 &$\pm$ 64  &(778) &     737 &$\pm$ 49  &(519)&     478 &$\pm$ 33  &(349) \\ 
\end{tabular}
}
\end{table}

\begin{table}[tbp]
  \centering
  \topcaption{
  Signal and background post-fit (pre-fit) yields in the \recolevel \njet bins. \label{tab:yields_njet}
  }
\cmsTable{  
\begin{tabular}{c r@{}l@{ }r r@{}l@{ }r r@{}l@{ }r r@{}l@{ }r r@{}l@{ }r}
          \multirow{2}{*}{Process} & \multicolumn{15}{c}{\recolevel \njet bin} \\ 
          & \multicolumn{3}{c}{0} & \multicolumn{3}{c}{1} & \multicolumn{3}{c}{2} & \multicolumn{3}{c}{3} &   \multicolumn{3}{c}{$\geq$ 4} \\

\hline
         Data    & \multicolumn{3}{c}{66263} & \multicolumn{3}{c}{42959} & \multicolumn{3}{c}{23027} & \multicolumn{3}{c}{8912} & \multicolumn{3}{c}{3765}  \\ [4pt] 
         H(125)  &   2186 &$\pm$ 92  &(2447)  &   1254 &$\pm$ 60  &(1165)  &    632 &$\pm$ 66  &(445)   &   178 &$\pm$ 48  &(109)   &    98 &$\pm$ 26 &(36) \\
All background   &  64085 &$\pm$ 463 &(63221) &  41650 &$\pm$ 374 &(43994) &  22367 &$\pm$ 344 &(22782) &  8735 &$\pm$ 182 &(8658)  &  3655 &$\pm$ 79 &(3822) \\ [4pt]
\tautau          &    740 &$\pm$ 41  &(520)   &    944 &$\pm$ 50  &(822)   &    688 &$\pm$ 99  &(301)   &   255 &$\pm$ 43  &(135)   &   100 &$\pm$ 50 &(70)\\
\ww              &  41058 &$\pm$ 360 &(38437) &  13190 &$\pm$ 252 &(15176) &   3402 &$\pm$ 222 &(4266)  &   698 &$\pm$ 125 &(966)   &     0 &$\pm$ 0  &(240) \\
\ttst            &  11125 &$\pm$ 144 &(11870) &  20891 &$\pm$ 179 &(21198) &  15788 &$\pm$ 214 &(15381) &  6853 &$\pm$ 110 &(6510)  &  3152 &$\pm$ 52 &(3031) \\
Nonprompt        &   6649 &$\pm$ 188 &(8999)  &   3436 &$\pm$ 149 &(4457)  &   1066 &$\pm$ 77  &(1792)  &   480 &$\pm$ 52  &(685)   &   254 &$\pm$ 30 &(357) \\
Other background &   4513 &$\pm$ 165 &(3394)  &   3189 &$\pm$ 139 &(2342)  &   1424 &$\pm$ 89  &(1043)  &   449 &$\pm$ 32  &(362)   &   149 &$\pm$ 12 &(124)\\
\end{tabular}
}
\end{table}
\section{Systematic uncertainties}
\label{sec:systematics}

The
experimental uncertainties mostly concern the accuracy in modeling the detector response in MC simulation, while the
theoretical uncertainties are more specific to individual signal and background processes. Because
signal extraction is performed using templates of \mllmtH distributions, the relevant effects
of the uncertainties are changes in the shapes and normalizations of the templates. In the
signal extraction fit, one continuous constrained nuisance parameter represents each such
change. The constraints are implemented through log-normal probability distribution functions, with
the nominal values of the nuisance parameters at zero and the widths given by the estimated sizes of the corresponding uncertainties.

Experimental uncertainties pertaining to all MC simulation samples, both signal and background, are
the uncertainties in trigger efficiency, lepton reconstruction and identification efficiencies,
lepton momentum scale, jet energy scale, and the uncertainty on \ptmiss arising from the momentum
scale of low \pt PF candidates not clustered into jets (unclustered energy). Uncertainties in lepton momentum and jet
energy scales also affect \ptmiss. Each of these uncertainties is represented by one independent nuisance parameter per
data set, effectively keeping the template variations for the three data sets in the simultaneous
fit uncorrelated. The uncertainty in {\cPqb} tagging efficiency, also included in this class of
uncertainties, is represented by seventeen nuisance parameters. Five of these nuisance parameters
relate to theoretical predictions of jet flavors involved in the measurement of the efficiency and
are thus common among the three data sets. The remaining twelve parameters, four per data set,
relate to statistical uncertainties in the samples used to measure the efficiency, and are
uncorrelated among the data sets~\cite{Sirunyan:2017ezt}.

Uncertainties in the trigger efficiency, and lepton reconstruction and identification efficiencies,
evaluated as functions of lepton \pt and $\eta$, cause variations in both the shape and the
normalization of the templates. The impacts on the template normalizations from the uncertainties in the trigger efficiency are less than
1\% overall, while the uncertainties in the reconstruction and identification efficiency cause shape and
normalization changes of $\approx$1\% for electrons and $\approx$2\% for muons. These uncertainties are dominated by the statistical fluctuations of the data set where they are measured, and are thus kept uncorrelated among the data sets.

Changes in the lepton momentum scale, the jet energy scale, and the unclustered energy scale all
cause migrations of simulated events between template bins and migration in and out of the
acceptance, which in turn cause changes in the shape and normalization of the templates. The impact
on the template normalization is $\approx$0.6--1.0\% in the electron momentum scale, 0.2\% in the
muon momentum scale, and 1--10\% in \ptmiss. For the changes in the jet energy scale, the impact on
the template normalization is $\approx$3 and 10\% in the \ptH and \njet measurements,
respectively. The latter has larger uncertainties because the jet energy scale directly affects the
number of events falling into different \recolevel \njet bins.

There are also experimental uncertainties in the estimation of the nonprompt lepton background. 
This background is affected by shape uncertainties arising from the dependence of the fake-lepton factors on the flavor composition of the jets misidentified as leptons. These shape uncertainties amount to $\approx$5--10\% (see Ref.~\cite{Sirunyan:2018egh} for details).
Additionally, a 30\% normalization uncertainty is assigned to the fit template for the nonprompt lepton
background from a closure test performed on simulation. 
Because these uncertainties depend on lepton reconstruction and
identification algorithms, which have differences among the three data sets, they are represented through
independent sets of nuisance parameters. Due to the difference in shape between the nonprompt lepton background and the other backgrounds and the signal, the normalization uncertainty is constrained post-fit to about 50\% of its pre-fit value. 

The uncertainties in the integrated luminosity are incorporated into the fit as changes in
normalization of the templates of the MC simulation samples, excluding the \ww, \ttst, and \tautau
samples. The total uncertainty in the CMS luminosity is 2.5, 2.3, and 2.5\% for the 2016, 2017,
and 2018 data sets,
respectively~\cite{CMS-PAS-LUM-17-001,CMS-PAS-LUM-17-004,CMS-PAS-LUM-18-002}. These evaluations are
partly independent, but also depend on inputs that are common among the three data sets. In total,
nine nuisance parameters are introduced to model the correlation in the uncertainties of the
integrated luminosity among the data sets.

Several theoretical uncertainties are relevant to all MC simulation samples. Uncertainties in this
category arise from the choice of the PDFs, missing higher-order corrections in the perturbative
expansion of the simulated cross sections, and modeling of the pileup. Template fluctuations due to
these uncertainties are controlled through nuisance parameters common to all three data sets. 

Since the changes in the shapes of the templates from the uncertainties in PDFs are found to be
small, only the normalization changes, both as cross section changes and acceptance changes, are
considered from this source. For the \ttst and \tautau events, while uncertainties in the overall
normalizations have no impact in the fit, uncertainties in PDFs give rise to respective 1\%
and 2\% uncertainties in the ratios of the predicted yields in the signal and the
control region.

Except for the ggF signal and \ww background processes, the estimated uncertainties from missing
higher-order corrections in the perturbative QCD expansion are given by the bin-by-bin difference
between the nominal and alternative templates, which are constructed from simulated events, where
renormalization and factorization scales are changed up and down by factors of two. Extreme
variations where one scale is scaled up and the other is scaled down are excluded. For the ggF
signal, the uncertainties are decomposed into several components, such as overall normalization and
event migrations between jet multiplicity bins~\cite{deFlorian:2016spz}. For the \ww background, the
higher-order corrections described in Section~\ref{sec:datasets} are modified by shifting the
renormalization and factorization scales and the jet veto threshold, where the latter determines the
scale below which QCD gluon radiation is resummed. The entire size of the electroweak corrections to the \ww process is taken as an uncertainty. For the uncertainties in both the PDF and
higher-order corrections, processes sharing similar QCD interactions are controlled through a common
nuisance parameter.

The uncertainty in the modeling of the pileup is assessed by changing the {\Pp{}\Pp} total inelastic cross
section of 69.2\unit{mb}~\cite{ATLAS:2016pu,Sirunyan:2018nqx} within a 5\% uncertainty, accounting for both the uncertainty in inelastic cross section measurement and the differences in primary vertex reconstruction efficiency between simulation and data.

Theoretical uncertainties in modeling the PS and UE primarily affect the jet multiplicity and are in
principle relevant to all MC simulated samples, but in practice have nonnegligible impacts on the
fit result only in the ggF and VBF signal samples and the quark-induced \ww background sample. The
uncertainty in the PS is evaluated by employing an alternative PS MC generator
(\HERWIGpp v2.7.1~\cite{Bahr:2008pv,Bellm:2015jjp}) for the simulation of the 2016 data set, and by
assigning PS variation weights computed in \PYTHIA~\cite{Mrenna:2016sih} to the simulated events for
the simulation of the 2017 and 2018 data sets. The UE uncertainty is evaluated by changing the fit
templates using MC simulation samples with UE tunes that are varied from the nominal tunes to cover
their uncertainties~\cite{Khachatryan:2015pea,Sirunyan:2019dfx}. For each of the PS and UE
uncertainties, changes in the 2017 and 2018 simulations are controlled through one nuisance
parameter, but the 2016 simulation uses an independent parameter.

In addition, there are theoretical systematic uncertainties specific to individual background
processes. The \ww background events have a 15\% uncertainty in the relative fraction of the
gluon-induced component~\cite{Caola:2016trd}. Similarly, the \ttst background events have an
uncertainty of 8\% in the fraction of the single top quark component. Also the \ttst background
sample considers the entire \pt correction weight (as mentioned in Section~\ref{sec:datasets}) as the
uncertainty in its \ttbar component. The \wgs process is assigned a 30\% uncertainty arising from the statistical precision of the trilepton control region used to estimate the scale factor assigned to this background process, as described in Section~\ref{sec:datasets}.

The theoretical uncertainties reflect those in the cross sections expected for signal processes, as
well as their template shapes. Because this analysis is a measurement of fiducial differential cross
sections, theoretical uncertainties in the fiducial cross section of each bin of 
\diffbasis must be excluded from the fits. This is achieved by keeping the normalizations of the
signal templates for individual \genlevel \diffbasis bins constant when changing the values of the nuisance
parameters corresponding to theoretical uncertainties.

It should be recognized that the use of regularization in signal extraction can introduce systematic
biases in the measured differential cross sections. In particular, by construction, a
discrepancy from the expectation in a single \diffbasis bin will be
suppressed if the neighboring bins do not exhibit discrepancies in the same direction. The scale of
possible regularization bias is measured from the results of the fit as outlined in
Ref.~\cite{Cowan}. 
In this method a toy data sample is created with signal yields corresponding to a statistical fluctuation around the best fit model. 
For each \diffbasis bin the difference in the number of events  
between the regularized fit result to the toy sample and the toy sample itself is taken as an indication of the scale of bias
introduced by regularization. 
These differences are then translated to estimates of
the bias in signal strengths through a multiplication by the rate of change of the extracted signal
strength modifiers, estimated by comparing the regularized fit result and the toy data sample. 
Estimated biases from regularization are separately reported in Section~\ref{sec:results} with the measured differential cross sections and other
uncertainties.
Unfolding bias has also been estimated as the difference between the true and fitted signal strength on an Asimov dataset constructed with either no VBF component or twice the expected VBF component. In this case the bias was smaller than the one estimated with the previously described method.

\section{Results}
\label{sec:results}

\begin{table}[b]
  \centering
  \topcaption{
    Observed signal strength modifiers and resulting cross sections in fiducial \ptH bins.
    The cross section values are the products of \ssm and the regularized $\mu$. The total uncertainty and the contributions by origin are given, where the
    contributions are statistical (stat), experimental excluding integrated luminosity (exp), theoretical related only to signal modeling (sig), to the
    background modeling (bkg), and integrated luminosity (lumi). Estimated biases in regularization are separately listed in the second from last column 
    and are not included in the total uncertainty.
  }
  \label{tab:obs_sigma_ptH}
  \cmsTable{
    \renewcommand{\arraystretch}{1.2}
    \begin{tabular}{ccccccccccc}
      \ptH & \ssm & \multirow{2}{*}{$\mu$} & \multicolumn{6}{c}{Regularized $\mu$} & \multirow{2}{*}{Bias} & \sobs \\
      ({\GeVns}) & ({\fbns}) & & Value & stat & exp & signal & bkg & lumi & & ({\fbns}) \\
      \hline
      0--20     & $27.45$ & $1.37\pm 0.30$ & $1.26\pm 0.27$ & $\pm 0.17$ & $\pm 0.19$ & $\pm 0.01$ & $\pm 0.10$ & $\pm 0.03$ & $+0.00$ & $34.6\pm 7.5$ \\
      20--45    & $24.76$ & $0.52\pm 0.42$ & $0.73\pm 0.36$ & $\pm 0.24$ & $\pm 0.25$ & $\pm 0.01$ & $\pm 0.10$ & $\pm 0.03$ & $-0.12$ & $18.2\pm 8.9$ \\
      45--80    & $15.28$ & $1.55\pm 0.41$ & $1.30\pm 0.33$ & $\pm 0.24$ & $\pm 0.20$ & $\pm 0.03$ & $\pm 0.09$ & $\pm 0.03$ & $-0.03$ & $19.9\pm 5.2$ \\
      80--120   & $7.72$ & $0.49\pm 0.52$ & $0.79\pm 0.42$ & $\pm 0.32$ & $\pm 0.25$ & $\pm 0.02$ & $\pm 0.08$ & $\pm 0.03$ & $-0.16$ & $6.1\pm 3.3$ \\
      120--200  & $5.26$ & $1.34_{-0.48}^{+0.51}$ & $1.14\pm 0.41$ & $\pm 0.29$ & $\pm 0.27$ & $\pm 0.04$ & $\pm 0.08$ & $\pm 0.03$ & $+0.11$ & $6.0\pm 2.2$ \\
      $>$200    & $2.05$ & $0.64_{-0.60}^{+0.63}$ & $0.73_{-0.57}^{+0.61}$ & $\pm 0.38$ & $\pm 0.42$ & $_{-0.03}^{+0.09}$ & $\pm 0.10$ & $\pm 0.03$ & $+0.19$ & $1.5\pm 1.2$ \\

    \end{tabular}
  }
\end{table}

\begin{table}[b]
  \centering
  \topcaption{
    Observed signal strength modifiers, uncertainties, and resulting cross sections in fiducial \njet bins.
    The cross section values are the products of \ssm and the unregularized $\mu$. 
    The uncertainties are separated by origin as in Table~\ref{tab:obs_sigma_ptH}.
  }
  \label{tab:obs_sigma_njet}
  \renewcommand{\arraystretch}{1.2}
  \begin{tabular}{ccccccccc}
    \multirow{2}{*}{\njet} & \ssm & \multicolumn{6}{c}{$\mu$} & \sobs \\
    & ({\fbns}) & Value & stat & exp & signal & bkg & lumi & ({\fbns}) \\
    \hline
    0  & $45.70$ & $0.88\pm 0.13$ & $\pm 0.06$ & $\pm 0.08$ & $\pm 0.01$ & $\pm 0.07$ & $\pm 0.03$ & $40.1\pm 6.0$ \\
    1  & $21.74$ & $1.06\pm 0.20$ & $\pm 0.12$ & $\pm 0.14$ & $\pm 0.01$ & $\pm 0.08$ & $\pm 0.03$ & $23.0\pm 4.6$ \\
    2  & $9.99$ & $1.50\pm 0.40$ & $_{-0.28}^{+0.25}$ & $\pm 0.28$ & $\pm 0.04$ & $\pm 0.11$ & $\pm 0.03$ & $15.0\pm 4.2$ \\
    3  & $3.26$ & $1.56_{-1.26}^{+1.35}$ & $_{-0.71}^{+0.89}$ & $_{-0.76}^{+0.84}$ & $_{-0.07}^{+0.17}$ & $_{-0.19}^{+0.29}$ & $_{-0.04}^{+0.07}$ & $5.1_{-4.1}^{+4.4}$ \\
    $\geq$ 4 & $1.83$ & $3.54_{-1.86}^{+2.05}$ & $_{-1.28}^{+1.10}$ & $_{-1.32}^{+1.28}$ & $_{-0.20}^{+0.40}$ & $_{-0.34}^{+0.38}$ & $_{-0.07}^{+0.10}$ & $6.5_{-3.4}^{+3.8}$ \\

  \end{tabular}
\end{table}
Tables~\ref{tab:obs_sigma_ptH} and \ref{tab:obs_sigma_njet} display the SM cross sections, observed
values of $\mu$, the uncertainties separated according to their origin, and the observed
cross sections. The contributions to the uncertainties are categorized as: statistical uncertainties in
the observed numbers of events; experimental uncertainties excluding those in the integrated
luminosity; theoretical uncertainties related only to signal modeling; other theoretical
uncertainties; and the uncertainties in the integrated luminosity. Table~\ref{tab:obs_sigma_ptH}
also shows the estimates of the regularization bias discussed at the end of
Section~\ref{sec:systematics}.

\begin{figure}[btp]
  \centering
  \includegraphics[width=0.45\textwidth]{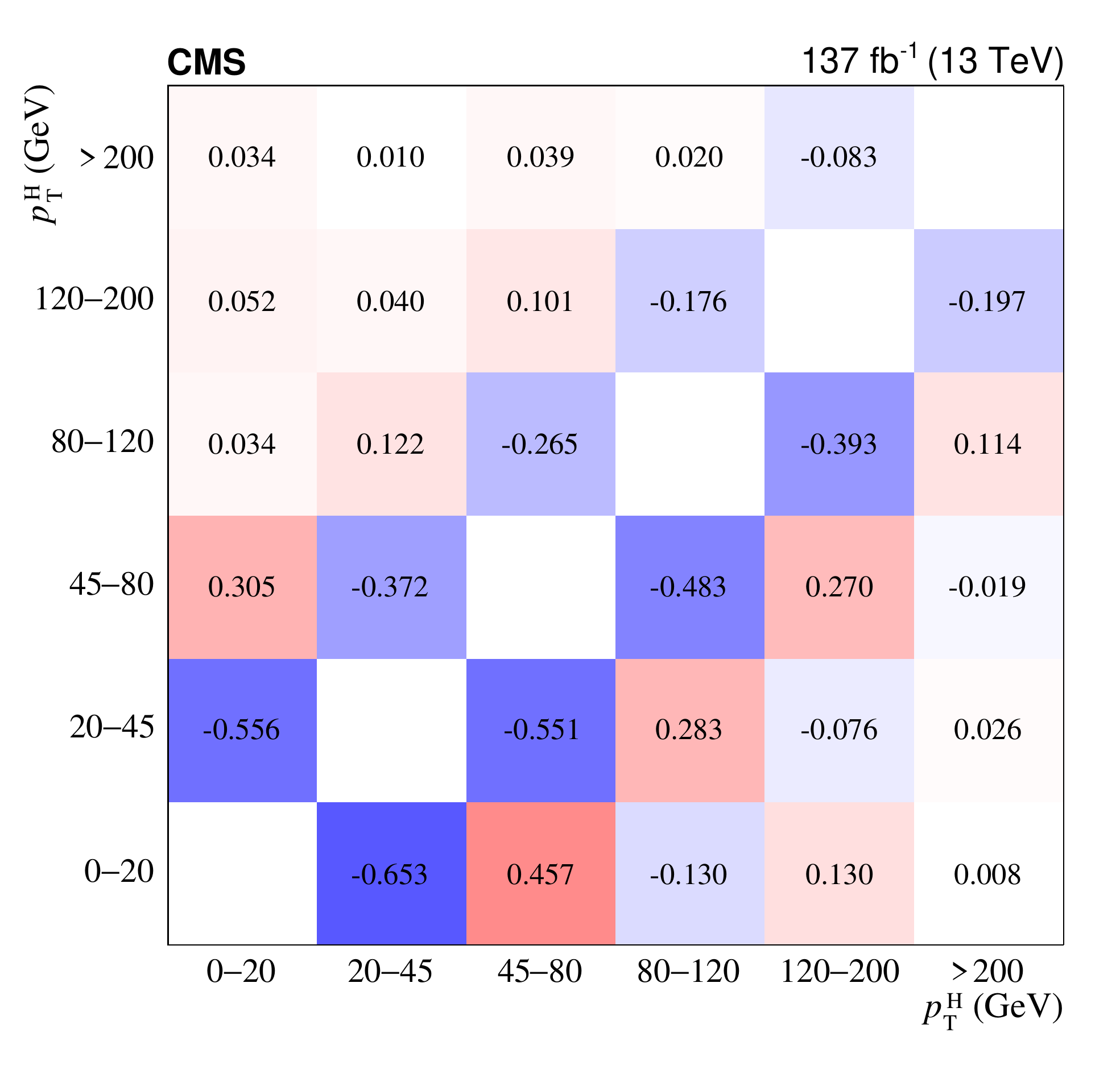}
  \includegraphics[width=0.45\textwidth]{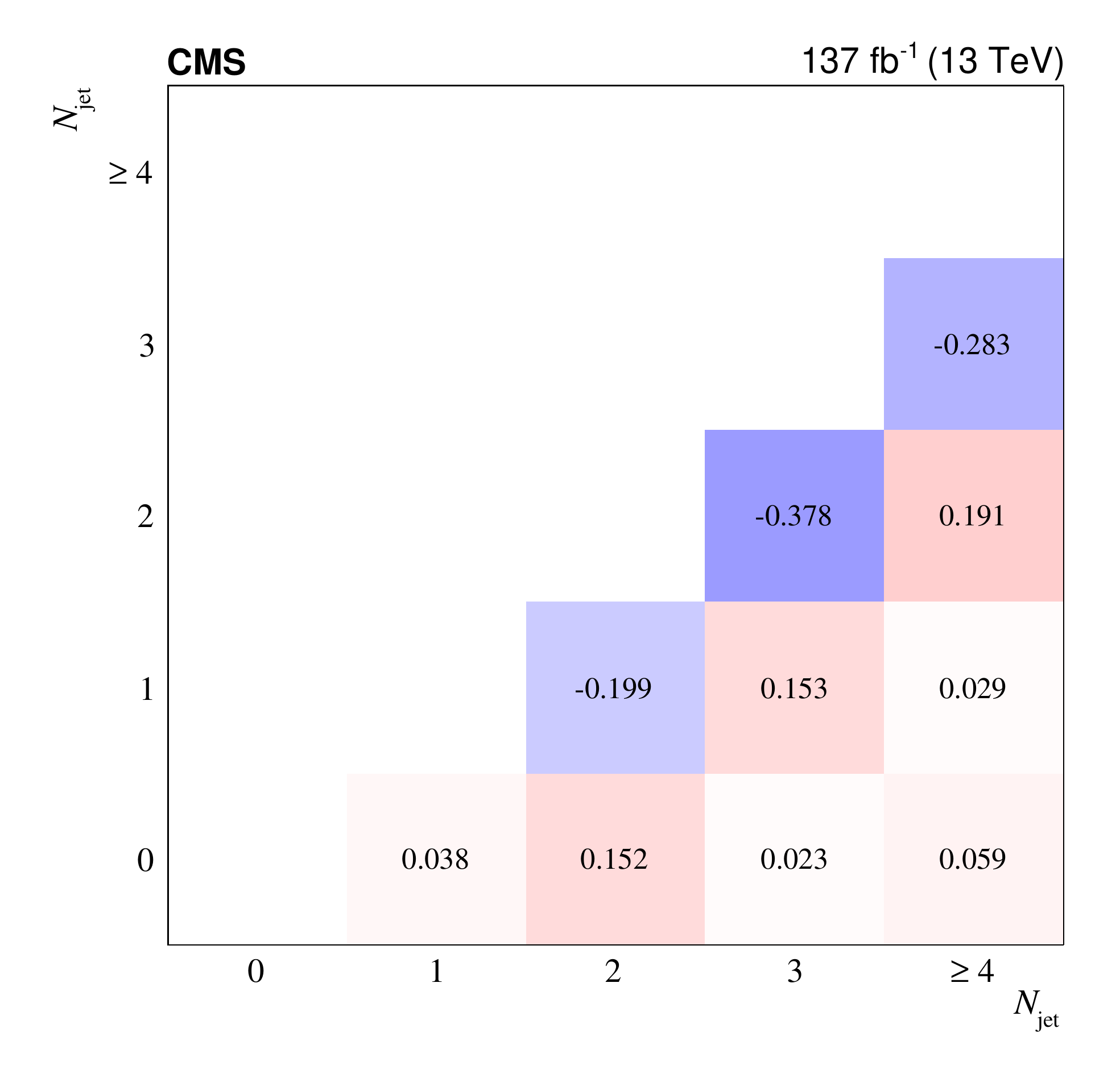}
  \caption{
    Correlation among the signal strength modifiers in bins of fiducial \ptH (left) and \njet (right). For the \ptH matrix, results of the regularized and unregularized fits are given above and below the diagonal.
  }
  \label{fig:mucorr}
\end{figure}

Correlations among the signal strength modifiers obtained from the fits are shown in
Fig.~\ref{fig:mucorr}. Because the \genlevel and \recolevel 
\diffbasis are not perfectly aligned, the signal template for a \genlevel bin has nonzero
contributions in neighboring  \recolevel bins. This misalignment induces negative
correlations between the signal strength modifiers of the nearest-neighbor bins in the fit, which
are indeed observed in the correlation matrices. Regularization counters this negative correlation,
as evident in the correlation matrix for the \ptH fit.

\begin{figure}[btp]
  \centering
  \includegraphics[width=0.45\textwidth]{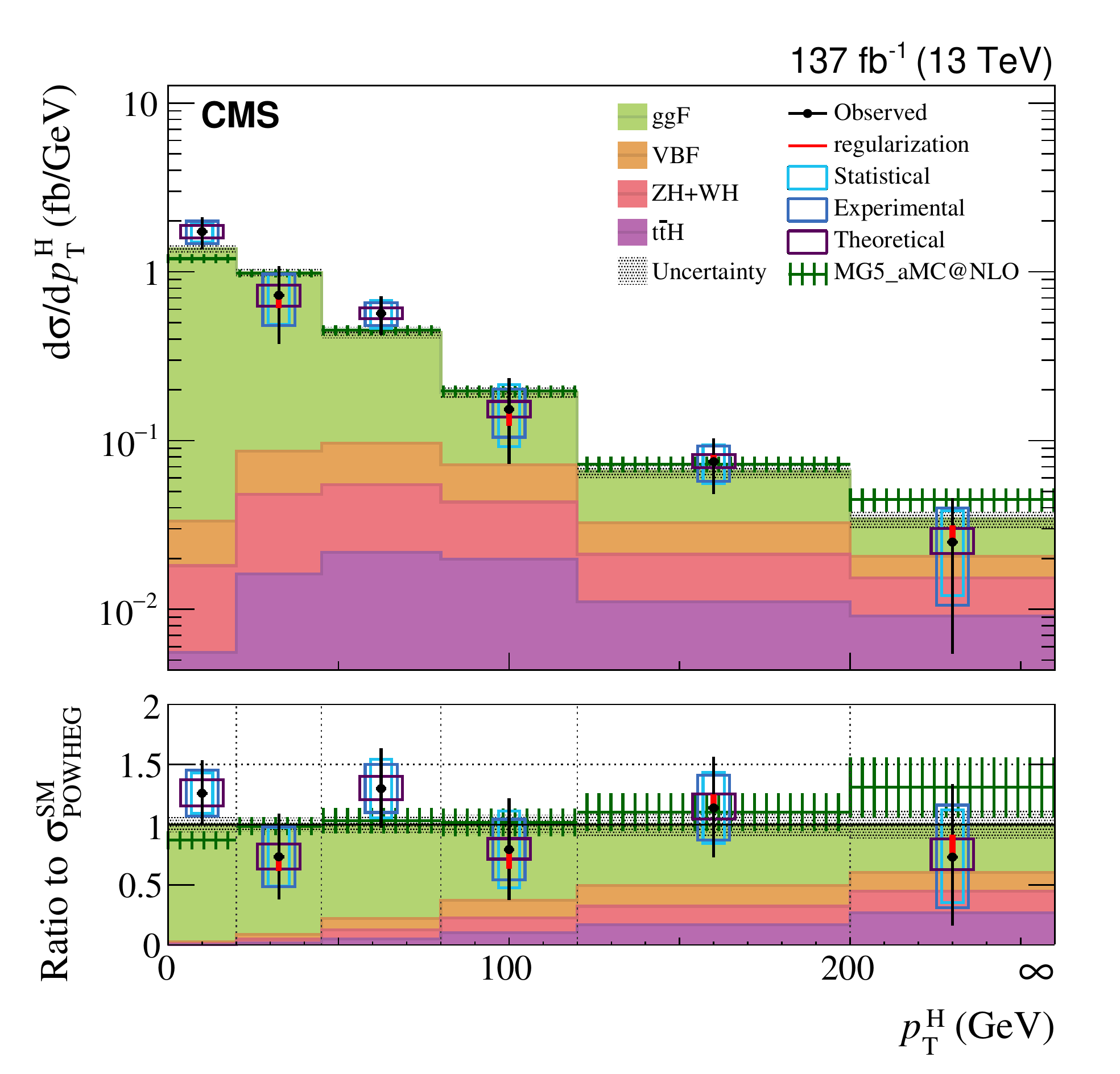}
  \includegraphics[width=0.45\textwidth]{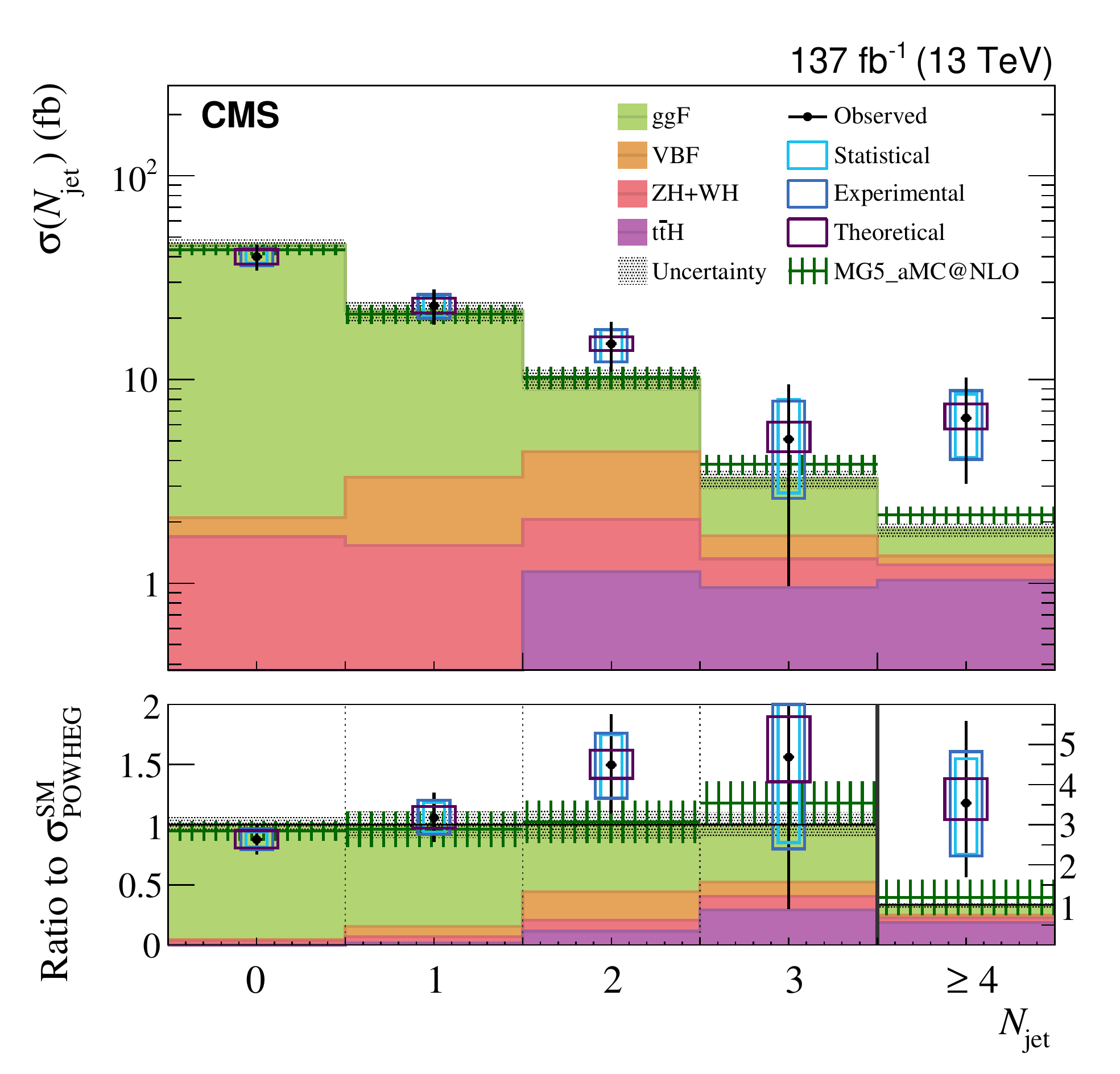}
  \caption{
    Observed fiducial cross sections in bins of \ptH (left) and \njet (right), overlaid with predictions from the nominal and alternative models for signal. The ggF and VBF samples are generated using \POWHEG in the nominal model and \MGvATNLO in the alternative model. The uncertainty bars on the observed cross sections represent the total uncertainty, with the statistical, experimental (including luminosity), and theoretical uncertainties also shown separately. The uncertainty bands on the theoretical predictions correspond to quadratic sums of renormalization- and factorization-scale uncertainties, PDF uncertainties, and statistical uncertainties of the simulation. The filled histograms in the ratio plots show the relative contributions of the Higgs boson production modes in each bin.
  }
  \label{fig:observed_sigma}
\end{figure}

The observed cross sections are compared with SM expectations in Fig.~\ref{fig:observed_sigma}. As
discussed in Section~\ref{sec:datasets}, all samples in the nominal signal model are generated using
\POWHEG, with the ggF component reweighted to match NNLO accuracy. Expectations from an
alternative signal model, where the \MGvATNLO generator is used for the ggF and VBF components but
the \VH and \ttH components are kept identical, are also overlaid in the figure.
The largest deviation from the SM prediction is observed in the $\geq$ 4 jet multiplicity bin and is 1.4 standard deviations.

In addition, the total fiducial cross section is extracted from a fit where the signal in Eq.~\ref{eqn:signal_term_original} is reformulated to
\begin{equation}
  s'_j(\mufid, \boldsymbol{\rho}; \boldsymbol{\theta}) = s_j(\mufid\boldsymbol{\rho}; \boldsymbol{\theta}) = \mufid \sum_{i} \left[ A_{ji}(\boldsymbol{\theta}) \rho_{i} L_{j} \sigmai \right],
\end{equation}
in which \mufid and all except one $\rho_{i}$ are free parameters. A specific $\rho_{k}$ depends on the other $\rho$ parameters via
\begin{equation}
  \rho_{k} = \frac{\ssm - \sum_{i\neq k} \rho_{i} \ssmi}{\ssm_{k}},
\end{equation}
fixing the sum $\sum_{i} \rho_{i} \ssmi$ to the total SM fiducial cross section \ssm, given in
Eq.~\ref{eqn:expected_sigma}. No regularization is applied for this fit. Through this reformulation,
anticorrelated components within uncertainties in \mui are absorbed into the sum
$\sum_{i}A_{ji}\rho_{i}\sigmai$, resulting in an uncertainty in \mufid that is smaller than the
quadratic sum of uncertainties in individual \mui that appear in Tables~\ref{tab:obs_sigma_ptH} and \ref{tab:obs_sigma_njet}.

The observed signal strength \mufid and cross section $\sfid = \mufid \ssm$ from the fit to the \ptH-binned combined data set, which has a smaller expected uncertainty than the fit to the \njet-binned combined data set, are

\begin{gather}
  \mufid  = 1.05\pm 0.12 \thickspace \Bigl(\pm 0.05\stat\pm 0.07\systexp\pm 0.01\systsignal\pm 0.07\systbkg\pm 0.03\lum\Bigr), \\
  \sfid   = 86.5\pm 9.5\unit{fb}.
\end{gather}
where (stat) refers to the statistical uncertainties (including the background normalizations extracted from control regions), (exp) to the experimental uncertainties excluding those in the integrated luminosity, (signal) to the theoretical uncertainties in modeling the signal, (bkg) to the remaining theoretical uncertainties, and (lumi) to the luminosity uncertainty.
Tabulated results are available in the HepData database~\cite{hepdata}.

\section{Summary}
\label{sec:summary}

Inclusive and differential fiducial cross sections for Higgs boson production have been measured
using \hwwenmn decays. The measurements were performed using {\Pp{}\Pp} collisions recorded by the
CMS detector at a center-of-mass energy of 13\TeV, corresponding to a total integrated luminosity of
137\fbinv. Differential cross sections as a function of the transverse momentum of the Higgs boson
and the number of associated jets produced are determined in a fiducial phase space that is matched
to the experimental kinematic acceptance. The cross sections are extracted through a simultaneous
fit to kinematic distributions of the signal candidate events categorized to maximize sensitivity to
Higgs boson production. The measurements are compared to standard model theoretical calculations using the
\POWHEG and \MGvATNLO generators. 
No significant deviation from the standard model expectations is observed. 
The integrated fiducial cross section is measured to be $86.5\pm 9.5\unit{fb}$,
consistent with the SM expectation of $82.5\pm 4.2\unit{fb}$.
These measurements were performed for the first time in the \hww decay channel at $\sqrt{s} = 13\TeV$ exploiting the full data sample available. The methods for the determination of the differential cross section have been updated significantly compared to the last report in the same channel at $\sqrt{s} = 8\TeV$, combining the signal extraction, unfolding, and regularization into a single simultaneous fit.

\begin{acknowledgments}

  We congratulate our colleagues in the CERN accelerator departments for the excellent performance of the LHC and thank the technical and administrative staffs at CERN and at other CMS institutes for their contributions to the success of the CMS effort. In addition, we gratefully acknowledge the computing centers and personnel of the Worldwide LHC Computing Grid for delivering so effectively the computing infrastructure essential to our analyses. Finally, we acknowledge the enduring support for the construction and operation of the LHC and the CMS detector provided by the following funding agencies: BMBWF and FWF (Austria); FNRS and FWO (Belgium); CNPq, CAPES, FAPERJ, FAPERGS, and FAPESP (Brazil); MES (Bulgaria); CERN; CAS, MoST, and NSFC (China); COLCIENCIAS (Colombia); MSES and CSF (Croatia); RIF (Cyprus); SENESCYT (Ecuador); MoER, ERC IUT, PUT and ERDF (Estonia); Academy of Finland, MEC, and HIP (Finland); CEA and CNRS/IN2P3 (France); BMBF, DFG, and HGF (Germany); GSRT (Greece); NKFIA (Hungary); DAE and DST (India); IPM (Iran); SFI (Ireland); INFN (Italy); MSIP and NRF (Republic of Korea); MES (Latvia); LAS (Lithuania); MOE and UM (Malaysia); BUAP, CINVESTAV, CONACYT, LNS, SEP, and UASLP-FAI (Mexico); MOS (Montenegro); MBIE (New Zealand); PAEC (Pakistan); MSHE and NSC (Poland); FCT (Portugal); JINR (Dubna); MON, RosAtom, RAS, RFBR, and NRC KI (Russia); MESTD (Serbia); SEIDI, CPAN, PCTI, and FEDER (Spain); MOSTR (Sri Lanka); Swiss Funding Agencies (Switzerland); MST (Taipei); ThEPCenter, IPST, STAR, and NSTDA (Thailand); TUBITAK and TAEK (Turkey); NASU (Ukraine); STFC (United Kingdom); DOE and NSF (USA).
  
  \hyphenation{Rachada-pisek} Individuals have received support from the Marie-Curie program and the European Research Council and Horizon 2020 Grant, contract Nos.\ 675440, 752730, and 765710 (European Union); the Leventis Foundation; the A.P.\ Sloan Foundation; the Alexander von Humboldt Foundation; the Belgian Federal Science Policy Office; the Fonds pour la Formation \`a la Recherche dans l'Industrie et dans l'Agriculture (FRIA-Belgium); the Agentschap voor Innovatie door Wetenschap en Technologie (IWT-Belgium); the F.R.S.-FNRS and FWO (Belgium) under the ``Excellence of Science -- EOS" -- be.h project n.\ 30820817; the Beijing Municipal Science \& Technology Commission, No. Z191100007219010; the Ministry of Education, Youth and Sports (MEYS) of the Czech Republic; the Deutsche Forschungsgemeinschaft (DFG) under Germany's Excellence Strategy -- EXC 2121 ``Quantum Universe" -- 390833306; the Lend\"ulet (``Momentum") Program and the J\'anos Bolyai Research Scholarship of the Hungarian Academy of Sciences, the New National Excellence Program \'UNKP, the NKFIA research grants 123842, 123959, 124845, 124850, 125105, 128713, 128786, and 129058 (Hungary); the Council of Science and Industrial Research, India; the HOMING PLUS program of the Foundation for Polish Science, cofinanced from European Union, Regional Development Fund, the Mobility Plus program of the Ministry of Science and Higher Education, the National Science Center (Poland), contracts Harmonia 2014/14/M/ST2/00428, Opus 2014/13/B/ST2/02543, 2014/15/B/ST2/03998, and 2015/19/B/ST2/02861, Sonata-bis 2012/07/E/ST2/01406; the National Priorities Research Program by Qatar National Research Fund; the Ministry of Science and Higher Education, project no. 02.a03.21.0005 (Russia); the Programa Estatal de Fomento de la Investigaci{\'o}n Cient{\'i}fica y T{\'e}cnica de Excelencia Mar\'{\i}a de Maeztu, grant MDM-2015-0509 and the Programa Severo Ochoa del Principado de Asturias; the Thalis and Aristeia programs cofinanced by EU-ESF and the Greek NSRF; the Rachadapisek Sompot Fund for Postdoctoral Fellowship, Chulalongkorn University and the Chulalongkorn Academic into Its 2nd Century Project Advancement Project (Thailand); the Kavli Foundation; the Nvidia Corporation; the SuperMicro Corporation; the Welch Foundation, contract C-1845; and the Weston Havens Foundation (USA).
\end{acknowledgments}

\bibliography{auto_generated}
\cleardoublepage \appendix\section{The CMS Collaboration \label{app:collab}}\begin{sloppypar}\hyphenpenalty=5000\widowpenalty=500\clubpenalty=5000\vskip\cmsinstskip
\textbf{Yerevan Physics Institute, Yerevan, Armenia}\\*[0pt]
A.M.~Sirunyan$^{\textrm{\dag}}$, A.~Tumasyan
\vskip\cmsinstskip
\textbf{Institut f\"{u}r Hochenergiephysik, Wien, Austria}\\*[0pt]
W.~Adam, F.~Ambrogi, T.~Bergauer, M.~Dragicevic, J.~Er\"{o}, A.~Escalante~Del~Valle, M.~Flechl, R.~Fr\"{u}hwirth\cmsAuthorMark{1}, M.~Jeitler\cmsAuthorMark{1}, N.~Krammer, I.~Kr\"{a}tschmer, D.~Liko, T.~Madlener, I.~Mikulec, N.~Rad, J.~Schieck\cmsAuthorMark{1}, R.~Sch\"{o}fbeck, M.~Spanring, W.~Waltenberger, C.-E.~Wulz\cmsAuthorMark{1}, M.~Zarucki
% ...
\vskip\cmsinstskip
\dag: Deceased\\
1:  Also at Vienna University of Technology, Vienna, Austria\\
% ...
\end{sloppypar}
\end{document}